\documentclass[12pt]{article}
\usepackage{amssymb}
\usepackage{latexsym}
\begin{document}
\def\r{\rho}\def\ro{\hat\r}
\def\b{\beta}
\def\Sth{S^{\rm thermo}}
\def\Dirr{\Delta_{\rm irr}}
\def\ox{\otimes}
\def\s{\sigma}\def\so{\hat\s}
\def\M{\mathcal{M}}
\def\tr{\mbox{tr}}

\title{On exact identity between thermodynamic and
informatic entropies in a unitary model of friction}
\author{Lajos Di\'osi\\
Research Institute for Particle and Nuclear Physics\\
H-1525 Budapest 114, POB 49, Hungary\\
Tova Feldmann and Ronnie Kosloff\\
Department of Physical Chemistry, Hebrew University\\
Jerusalem 91904, Israel}
\maketitle

\begin{abstract}
An elementary collision model of a molecular reservoir is considered upon which an external field 
is applied and the work is dissipated into heat. To realize macroscopic irreversibility at the
microscopic level, we introduce a ``graceful'' irreversible map which randomly mixes the identities 
of the molecules. This map is expected to generate informatic entropy exactly equal to the independently
calculable irreversible thermodynamic entropy. 
\end{abstract}

\section{Introduction}
The thermodynamic entropy $\Sth_R$ of a given equilibrium reservoir is equal to the informatic 
(Shannon--von Neumann) entropy 
\begin{equation}
S[\r_R]\equiv -\tr(\r_R\log\r_R). 
\end{equation} 
The $\r_R$ is the corresponding Gibbs canonical state of the reservoir, 
taken in the infinite volume (thermodynamic) limit.
If we switch on a certain external macroscopic field to perform work on the reservoir 
then the state $\r_R(t)$ starts to evolve reversibly while, from the thermodynamic viewpoint, 
part of the work will be dissipated into heat in the reservoir. There is a common expectation
that the irreversible thermodynamic entropy production also equals the change of the informatic entropy 
of the reservoir:
\begin{equation}\label{Deltairr}
\Dirr\Sth_R = S[\r_R(t+\Delta t)]-S[\r_R(t)].
\end{equation}
As it is well-known this equation does not hold because $\r_R(t)$ evolves unitarily and
the r.h.s. is always zero. This is the notorious conflict between the reversibility of the
reservoir's microscopic dynamics and the macroscopically observed irreversible dissipation.
In our case this dissipation is ``friction'' against the external field. Various resolutions of this
contradiction between the microscopic and macroscopic theories are possible.
None of them is general and none of them is able to prove the exact
identity of the thermodynamic and informatic entropy productions. Obviously, any 
resolution must {\it impose some deliberate irreversibility} on the unitary dynamics. For
a systematic study, we formalize this irreversibility as a completely positive map $\M$ 
applied to the reservoir state repeatedly. Then we expect that
\begin{equation}\label{DeltairrM}
\Dirr\Sth_R = S[\M\r_R(t+\Delta t)]-S[\r_R(t)]
\end{equation}
holds exactly for microscopically short intervals $\Delta t$. The map $\M$ we are looking
for must be {\it graceful}. It should not alter the macroscopic reaction (e.g. the friction force) 
of the reservoir on the field. Moreover, it should be such that it conserve the free
dynamics of the reservoir:
\begin{equation}\label{mapHrho}
\M[H_R,\r_R(t+\Delta t)]=[H_R,\M\r_R(t+\Delta t)],
\end{equation}
where $H_R$ is the reservoir's Hamitonian.

In our work we construct an elementary reservoir model composed of distinguishable molecules
and postulate a
discrete reversible (unitary) collision dynamics for its interaction with the external field.
In this model the thermodynamic entropy production $\Dirr\Sth$ is exactly calculable.
Then a natural projection $\M$ is found that expected to increase the information entropy of the 
reservoir by $\Dirr\Sth$ exactly. The map $\M$ symmetrizes the state $\r_R$ for all distinguishable 
molecules, i.e., it mixes the ones which collided with the external field among all molecules throughout 
the reservoir. The identity (\ref{DeltairrM}) relies upon a novel conjecture concerning the increase 
of the informatic entropy under the projection $\M$. We have heuristically proved the theorem for the 
classical case. 

\section{The reservoir, collisions}
Assume a single molecule with a Hamiltonian $H$ in the Gibbs-equilibrium state 
$\r$ at inverse temperature $\b$:
\begin{equation}\label{Gibbs}
\r=\frac{1}{Z(\b)}e^{-\b H}.
\end{equation}
Let the molecule interact with a certain external field and
let us describe the interaction by the following ``collision'':
\begin{equation}\label{U}
\r\rightarrow U\r U^\dagger\equiv\s,~~~~~[U,H]\neq 0,
\end{equation}
where $U$ is unitary. 
Suppose our reservoir is formed by the ideal gas of $n$ such {\it distinguishable} molecules
in equilibrium:
\begin{equation}
\r_R = \r\ox\r\ox\dots\r\equiv\r^{\ox n},
\end{equation}
with the Hamiltonian
\begin{equation}
H_R=H\ox I^{\ox(n-1)} + I\ox H\ox I^{\ox(n-2)} +I^{\ox(n-1)}\ox H,
\end{equation}
and its thermodynamic limit is $n\rightarrow\infty$. Let us assume that
the molecules will reversibly collide with the field in succession, see Eq.(\ref{U}).
Without restricting the generality, we can assume that the 1st molecule
collides first, the 2nd collides second, {\it e.t.c.}:
\begin{equation}
\r^{\ox n}\rightarrow\s\ox\r^{\ox(n-1)}\rightarrow\s\ox\s\ox\r^{\ox(n-2)}\rightarrow\dots
\end{equation}
Since $S[\s]=S[\r]$,
the informatic entropy of the reservoir has not 
changed at all in the above reversible collisions:
\begin{equation}
S[\r^{\ox n}]=S[\s\ox\r^{\ox(n-1)}]=S[\s\ox\s\ox\r^{\ox(n-2)}]=\dots=
nS[\r]=nS[\s].
\end{equation}

For simplicity, we consider the 1st collision and implement our idea for it.
If we calculate the mean energy transfer from the field to the molecule
we obtain:
\begin{equation}\label{DeltaE}
\Delta E\equiv\tr(\s H)-\tr(\r H)=\frac{\tr(\s\log\s)-\tr(\s\log\r)}{\b}~~\rangle~~0.
\end{equation}
The proof is straightforward if we express $\log\r=-\b H -\log Z$ from (\ref{Gibbs}),
substitute it, and observe that $\tr(\r\log\r)=\tr(\s\log\s)$.
The r.h.s. above is proportional to the informatic relative entropy $S[\s|\r]$ of the
post- and pre-collision states, which is always positive. In a reversible process, a 
Gibbs-state will always {\it absorb} energy from the field.
If the molecule is part of a reservoir and the field interacts with many
molecules in succession then, thermodynamically, we expect that the above 
mean energy is dissipated to the reservoir. Therefore the average thermodynamic
entropy production per collision is, on thermodynamic grounds:
\begin{equation} 
\Dirr\Sth=\b\Delta E=S[\s|\r].
\end{equation}
We see that $\Dirr\Sth$ is always positive since $S[\s|\r]$ is
always positive if $\s\neq\r$.
We emphasize that the above identity 
is part of common knowledge and is different from our target identity.
The relative entropy which is microscopic
plays an auxiliary role in the calculations. We are looking for the informatic 
entropy production whose value shall coincide with the relative entropy. 
Our target identity is:
\begin{equation} 
\Dirr\Sth=\Delta S_R~,
\end{equation}
where $\Delta S_R$ is the {\it increase} of the informatic entropy between
the post- and pre-collision states of the {\it whole reservoir}. We already 
noticed that $\Delta S_R$ would remain zero after the 1st unitary collision.
The point is that we have to define a new post-collision state of the reservoir, which 
differs from $\s\ox\r^{\ox(n-1)}$.

\section{The graceful irreversible map}
Now we have to {\it postulate} a (completely) positive
map $\M$ which is irreversible and {\it graceful}:
\begin{equation}
\s\ox\r^{\ox(n-1)}\rightarrow\M\left(\s\ox\r^{\ox(n-1)}\right).
\end{equation}
The map should correlate the molecules. Single-molecule maps can increase
the informatic entropy but they would not produce the requested value $S[\s|\r]$.
We need a map which does only smear out information whose loss is heuristically justifiable
in a molecular reservoir.
Let the map $\M$ be the total randomization over the identity of the $n$ molecules.
In our case:
\begin{equation}
\M\left(\s\ox\r^{\ox(n-1)}\right)
=\frac{\s\ox\r^{\ox(n-1)}+\r\ox\s\ox\r^{\ox(n-2)}+\dots
                       +\r^{\ox(n-1)}\ox\s}{n}~.
\end{equation}
It is clear that this post-collision map is irreversible and increases the informatic
entropy of the reservoir. Moreover, it is consistent with Eq.(\ref{mapHrho}) in
conserving the energy.

In the Appendix we are going to illustrate for the
classical case that the informatic entropy increases by the relative entropy of the
post- and pre-collision states. According to this novel mathematical conjecture:
\begin{equation}
\Delta S_R\equiv \lim_{n=\infty}
\left( S[\M(\s\ox\r^{\ox(n-1)})]-S[\s\ox\r^{\ox(n-1)}] \right) =S[\s|\r].
\end{equation}
This would confirm the central physical identity (\ref{DeltairrM}). 
This says that our model yields exact identity between the independently
defined thermodynamic and informatic entropies. The key element of the model is
a graceful irreversible map postulated after each reversible interaction of the
field with a molecule. 

\section{Discussion}
Realistic microscopic models of dissipation are not calculable exactly. Note that both 
the irreversible thermodynamic $\Dirr\Sth$ and the informatic $\Delta S_R$ entropies 
should be calculable independently and exactly. Our model is less
realistic while it allows exact calculation and confirmation of  
$\Dirr\Sth=\Delta S_R$. Our elementary
model can be improved or varied in certain ways. For instance, the map $\M$  
does not need to be repeated after each collision, it may be applied after several
collisions --- the main result remains. The collision dynamics is, however,
essential. If we resolved it into a smooth Hamiltonian evolution the model would
become fundamentally sensitive to the repetition frequency of the irreversible map
$\M$.  

In the spirit of the present work, a particular version of the model appeared already 
in \cite{Dio} as the model of classical mechanical friction. The relationship of the 
map $\M$ to the Gibbs principle of molecular undistinguishability has been discussed 
although not fully persued. This is probably a deeper issue and subject of future
research. 

Analysis of friction phenomena in a reduced single molecule description can be based
on a two step cyclic process \cite{Fel}. The first is identical to Eq.(\ref{U}). The
second step is an irreversible  map generated by the phenomenological Lindblad operator 
which restores the system to its initial state. The cyclic requirement means that all 
entropy has to be generated in the reservoir. Unlike the present model, in the Lindblad 
formulation there is no explicit description of the graceful map that generates the 
desired entropy production.

\section{Asher}
Asher has strived for internal consistency between independent but related branches of 
physics. This outlook is reflected in his Book \cite{Per93} where he has linked numerous 
thermodynamical aspects to quantum theory and vice versa. 

Following von Neumann, Asher proves that the equilibrium quantum informatic entropy 
is genuine thermodynamic entropy (p. 270). One reads the derivation of 
a version of our Eq.(\ref{DeltaE}), see p. 269, Eq.(9.24). One also learns the 
heuristic derivation of the 
Shannon informatic entropy (p. 260) we exploited in the Appendix. Asher invoked 
thermodynamical argument particular in a debate on hypothetic nonlinear Schr\"odinger
equations, p. 278, see also \cite{Per89}. The power of thermodynamic arguments may show itself 
in our present work: the novel mixing entropy conjecture (Appendix) for quantum entropies   
has so far not been proven but it is strongly supported by the physical wisdom that the
thermodynamic and informatic entropy productions should coincide.

{\it Acknowledgment.\/}
We thank the Israel Science Foundation and the Hungarian 
OTKA Grant 49384.

\appendix
\section{Conjecture on entropy of mixing}
We consider $n+1$ uncorrelated identical systems where $n$ systems
have the same state $\r$ and a single system has a certain different
state $\s$, where both $\r$ and $\s$ can be given arbitrarily.
Consider the following uncorrelated composite state: 
\begin{equation}
\s\ox\r^{\ox n}.
\end{equation}
Then we assume that the single different system ``looses'' its identity
among the other components, i.e., it becomes totally mixed with them.
Let the resulting state be this:
\begin{equation}\label{R}
R=\frac{\s\ox\r^{\ox n}+\r\ox\s\ox\r^{\ox(n-1)}+\dots
                       +\r^{\ox n}\ox\s}{n+1}~.
\end{equation}
We define the entropy of mixing:
\begin{equation}\label{Smixn}
S^{\rm mix}[\s\vert\r~;n]=S[R]-nS[\r]-S[\s].
\end{equation}
We conjecture that in the limit $n\rightarrow\infty$ the entropy of mixing is identical 
with the relative entropy of $\s$ with respect to $\r$:
\begin{equation}\label{Smixinf}
S^{\rm mix}[\s\vert\r~;\infty]=-\tr(\s\log\r)-S[\s]\equiv S[\s\vert\r]
\end{equation}
The proof of the general quantum case is missing. We present a heuristic
proof for the classical case $[\s,\r]=0$.

{\it Heuristic proof -- classical case.\/}
We can directly estimate the increase of Shannon entropy caused by
mixing. Let us consider a discrete $d$-state system and its diagonal density 
matrices with elements $\r_{ab}=\r_a\delta_{ab}$ and  
$\s_{ab}=\s_a\delta_{ab}$, where $a,b=1,2,\dots,d$. If $n$ is large,
we can ignore all but the statistically relevant terms in $\ro^{\ox n}$.
Look at the components
\begin{equation}
\r_{a_1}\r_{a_2}\dots\r_{a_n}~,
\end{equation}
and consider the string 
\begin{equation}\label{a1a2a3}
a_1a_2a_3\dots a_n
\end{equation}
of labels.
In the statistically relevant terms, the multiplicity of a given index 
$a$ must approximately be $n\r_a$ for each $a=1,2,\dots,d$. Therefore,
each relevant term has the same weight and the total number of relevant
terms can be estimated combinatorically as
\begin{equation}\label{Shannon}
\frac{n!}{(n\r_1)!(n\r_2)!\dots(n\r_d)!}\sim 2^{nS[\r]}~,
\end{equation}
which is the number of all different orderings within the string (\ref{a1a2a3}).
Now, take the single state described by $\s_a$, insert the latter into
the string at random and try to calculate the expected increase of the
number of relevant configurations. Assume temporarily that $\s_a=\delta_{a1}$.
Then the above number of combinations acquires a factor
\begin{equation}\label{factor}
\frac{n+1}{n\r_1+1}\sim\frac{1}{\r_1}~.
\end{equation}
Eq.(\ref{factor}) means that the entropy increases by $-\log\r_1$. 
Generalizing this result, we can estimate the average increase of the 
entropy for any distribution $\s_a$:
\begin{equation}
-\sum_a \s_a\log\r_a -S[\s]=S[\s|\r],
\end{equation}
which is the classical special case of our conjecture (\ref{Smixinf}).

\end{document}